\documentclass[12pt]{article}
\usepackage{graphicx}

\def\Title#1{\begin{center} {\Large #1 } \end{center}}
\def\Author#1{\begin{center}{ \sc #1} \end{center}}
\def\Address#1{\begin{center}{ \it #1} \end{center}}

\newenvironment{Abstract}{\begin{quotation}  }{\end{quotation}}

\def\beq{\begin{equation}}
\def\eeq#1{\label{#1}\end{equation}}
\def\eeqn{\end{equation}}

\def\beqa{\begin{eqnarray}}
\def\eeqa#1{\label{#1}\end{eqnarray}}
\def\eeqan{\end{eqnarray}}

\let\bar=\overbar

\def\Dslash{\not{\hbox{\kern-4pt $D$}}}
\def\dslash{\not{\hbox{\kern-2pt $\del$}}}

\def\msb{{\bar{\ssstyle M \kern -1pt S}}}

\usepackage{amsmath,amssymb,amsfonts}
\usepackage{graphicx,epsfig, subfigure}
\usepackage{hyperref}
\usepackage{color}
\usepackage{slashed}
\usepackage{epigraph}
\usepackage{pifont}
\usepackage{hyperref}
\usepackage{slashed}
\usepackage{epigraph}
\usepackage{authblk}

\newcommand{\bear}{\begin{array}}
\newcommand{\ear}{\end{array}}

\def\OMIT#1{{}}
\newcommand{\lsim}{\mathrel{\rlap{\lower4pt\hbox{\hskip1pt$\sim$}}
    \raise1pt\hbox{$<$}}}         %less than or approx. symbol
\newcommand{\gsim}{\mathrel{\rlap{\lower4pt\hbox{\hskip1pt$\sim$}}
    \raise1pt\hbox{$>$}}}         %greater than or approx. symbol

\newcommand{\ba}{\begin{eqnarray}}
\newcommand{\ea}{\end{eqnarray}}

\begin{document}
%\maketitle
\Title{Machine Learning for New Physics Searches}
\Author{Raffaele Tito D'Agnolo}
\Address{SLAC National Accelerator Laboratory, Menlo Park, CA  94025, USA}
\begin{Abstract}
This is the summary of a talk based on~\cite{DAgnolo:2018cun}. I briefly introduce neural networks and then discuss a new technique for model-independent new physics searches in background-dominated datasets.
\end{Abstract}

\begin{quotation}
\begin{center}
PRESENTED AT \\
CIPANP 2018: May 29 - June 3, 2018 Palm Springs, CA
\end{center}
\end{quotation}

\section{Introduction}
Today in particle physics and cosmology, we have collected large datasets that are accurately described by the models developed in the last decades. We have theoretical reasons to expect new particles and new dynamics beyond these reference models, especially in particle physics, but no consensus on how it should manifest itself. In this context, a tool to look for discrepancies between the datasets at our disposal and the reference models independently of our favorite new physics model, would be extremely valuable. In this document I present a new application of neural networks to physics that realizes this objective. We first proposed it in~\cite{DAgnolo:2018cun}.

Machine learning techniques have been steadily developing, obtaining remarkable successes in the last two decades. %Most of the theoretical ideas underlying machine learning and its various close relatives are applications of statistics. The basic principles behind this data analysis revolution were laid out in the 50s, with some core techniques being as old as 1847~\cite{S2010}. However the field recently outgrew its theoretical foundations, accumulating striking successes in a wealth of practical applications that are still far from being fully understood. This was possible thanks to recent increases in computing power and a few algorithmic breakthroughs.
Today their applications to fundamental research and commercial markets are countless. The most relevant physics examples concern large datasets in astrophysics, astronomy and collider physics. Large Hadron Collider (LHC) experiments are already extensively using these techniques to improve their particle identification capabilities and the sensitivity of their analyses. 
 %~\cite{Ball2009,Lahav:1995wv,Rajpaul:2012wu,2006PABei..24..285L} and collider physics%~\cite{Cranmer:2016swd,Baldi:2016fzo,Cranmer:2015bka,Chang:2017kvc,Cohen:2017exh,Komiske:2017ubm, deOliveira:2015xxd, Schwartzman:2016jqu, Kagan:2016wnu, Larkoski:2017jix, Louppe:2017ipp, Shimmin:2017mfk, Baldi:2016fql, Guest:2016iqz, Almeida:2015jua, Barnard:2016qma, Kasieczka:2017nvn, Butter:2017cot, Datta:2017rhs, Datta:2017lxt, Fraser:2018ieu, Andreassen:2018apy, Macaluso:2018tck, ATLAS:2017jiz, CMS-DP-2017-013, ATL-PHYS-PUB-2017-013, ATL-PHYS-PUB-2017-004, CMS-DP-2017-005, ATL-PHYS-PUB-2017-003, ATL-PHYS-PUB-2017-017, CMS-DP-2017-027, Baldi:2016fzo, Chang:2017kvc, Cohen:2017exh, Brehmer:2018eca, Brehmer:2018kdj, Brehmer:2018hga, Roxlo:2018adx, Collins:2018epr}. 
The typical application consists in optimizing the discrimination between events generated by different physical processes in multivariate samples. For instance distinguishing between gluon production and quark production at hadron colliders~\cite{Metodiev:2017vrx,Komiske:2018oaa,Dery:2017fap}. %To my knowledge there are no applications of machine learning to particle physics that significantly deviate from this paradigm. 
Starting with two or more types of events one tries to optimally separate them in a multidimensional parameter space. Essentially the neural network is used to find the non-linear combinations of the input variables that best discriminate between known classes of events. 

In this document I discuss a qualitatively new application of machine learning techniques to searches for new physics in large datasets. We directly exploit the properties of neural networks as universal approximants~\cite{Function1989, Hornik1991,Kreinovich1991, Haykin1998}. The main difference with respect to known applications in physics is that we will assume to know nothing about the form of the signal that we are looking for. We are not trying to discriminate between two known classes of events, but rather check if a dataset follows a given reference model. Effectively we are not making any assumption on the alternative model that best describes the data. The reader familiar with statics can already predict the onset of the look-elsewhere effect~\cite{Cowan:1998ji}, but we have a straightforward way to quantify it correctly as discussed in Section~\ref{sec:algorithm}.

As is the case in cosmology and particle physics, imagine that we have a reference model that correctly predicts the bulk of our observed events. Then we would like to look for small deviations on top of a largely background-dominated dataset. The form of these deviations is not determined a priori and the task of our machine learning algorithm is to find the most promising anomaly in the dataset. As a byproduct of our effort the network is going to output the ratio between the probability distribution function of the data and that of the reference model. Therefore if any anomaly is found we can trace back its origin in a fully transparent way, finding a description in terms of the simple observables given as inputs to the neural network.

In the next Section I briefly review the basic concepts of machine learning needed to understand our anomaly detection strategy, starting from what is a neural network from the point of view of a theoretical physicist. In Section~\ref{sec:algorithm} I describe our proposed technique for new physics searches and its possible applications.

\section{Neural Networks for the Busy Theoretical Physicist}\label{sec:nn}
A neural network is a way to parametrize a set of functions built out of nested units 
\ba
f_{NN}(\vec x)=f\circ g\circ h \circ ... \circ \vec x =f(g(h( ... (\vec x))))\, .
\ea
The units that form this basis can be of two kinds. Linear transformations
\ba
h(\vec x) = \vec w \cdot \vec x + b\, , \label{eq:linear}
\ea
that depend on the free parameters $w$, called weights, and $b$, called biases, and non-linear functions that are fixed (do not depend on extra free parameters) and are applied to each element of their input 
\ba
g(\vec x) = \left[g(x_1), g(x_2), ... , g(x_N)\right]\, . \label{eq:nonlinear}
\ea
The elements of the set of functions described by the network are spanned by weights and biases. The process of finding the optimal values of these free parameters goes under the name of training or learning and is very similar to old-fashioned fitting. I describe the procedure in some more detail at the end of this Section. 

For simplicity in Eq.~(\ref{eq:linear}) I have written a linear transformation that outputs a single number, but obvious higher-dimensional generalizations are possible. Examples of non-linear functions that have found successful applications in the past are the logistic sigmoid $\sigma(x)=1/(1+e^x)$, the Rectified Linear Unit (ReLU), $R(x) = x \theta(x)$ and the hyperbolic tangent. They all have several qualitative features in common. They saturate for large values (negative, positive or both) of their argument, have a turn-on and are monotonically increasing. In a few paragraphs I will present a heuristic argument that shows why these characteristics are useful if one needs to approximate an arbitrary non-linear function without introducing too many free parameters. 

\begin{figure}[!t]
\begin{center}
\includegraphics[width=0.6\textwidth]{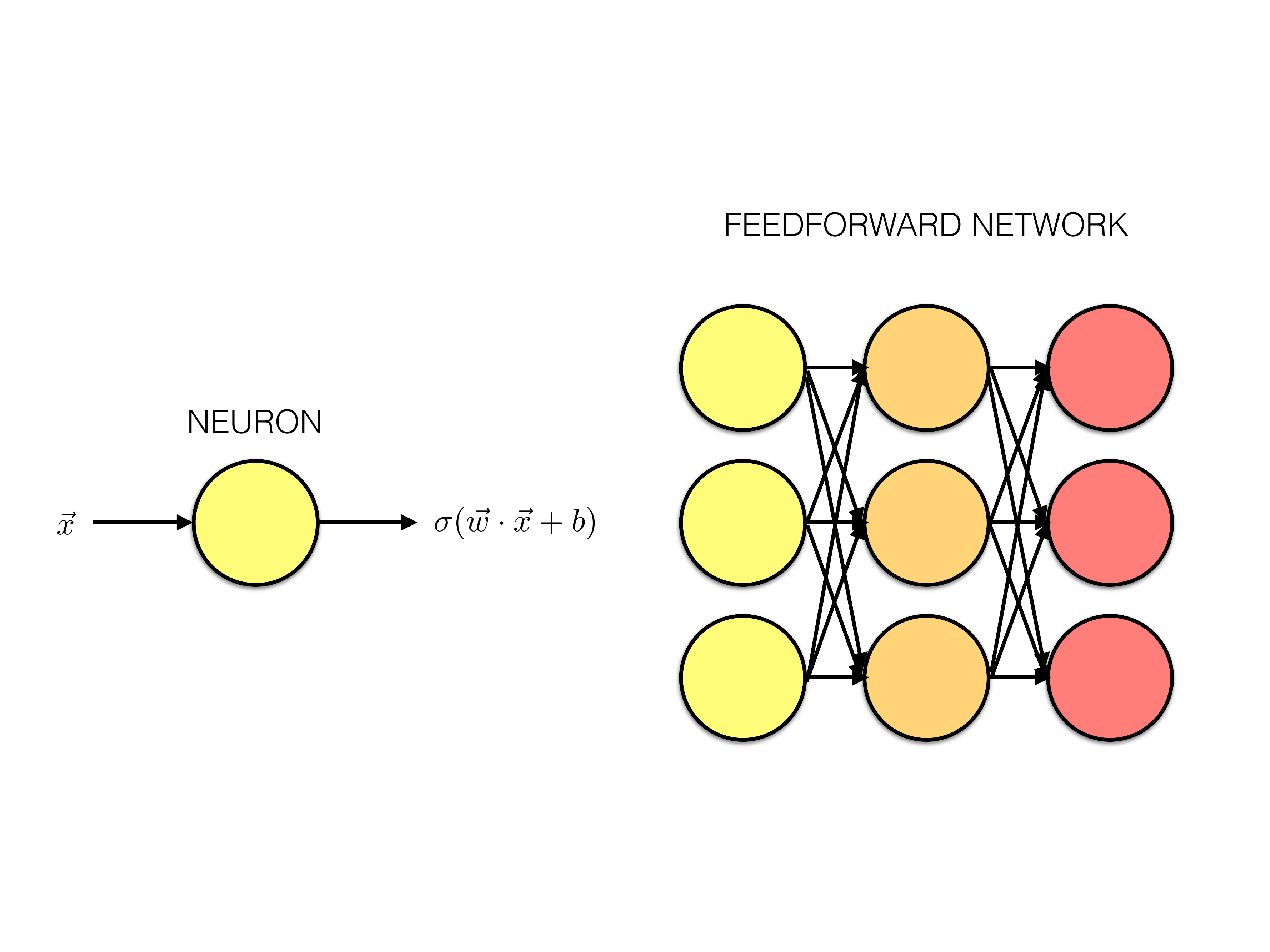}
\caption{Left: Schematic representation of a neuron with a logistic sigmoid activation function. Right: Schematic representation of a fully connected feedforward neural network. The input and output layers of the network are not shown.}
\label{fig:network}
\end{center}
\end{figure}

The procedure of nesting these classes of functions allows for a vast range of possibilities. Here we consider one of the simplests, which is more clearly explained in terms of ``neurons". I take a neuron to be a transformation that applies to its input the function in (\ref{eq:linear}) followed by (\ref{eq:nonlinear}). So, using a logistic sigmoid as an example, a neuron corresponds to the function
\ba
f_N(\vec x | \vec w, b) = \sigma(\vec w \cdot x +b)\, . \label{eq:neurons}
\ea
In introductory machine learning texts you will often find the neuron to be defined just as the linear transformation with the non-linear piece being called an activation function. Terminology aside we can now build a neural network by linking neurons together. A typical example is shown in Figure~\ref{fig:network}. Neurons are organized in layers. The first layer takes as input elements of the dataset of interest. Subsequent inner layers take as input the output of the previous layer. 

If we send the output of all the neurons in layer $n$ to all the neurons in layer $n+1$ we are building a fully connected network. Letting information flow only in one direction is known as building a feedforward network. So Figure~\ref{fig:network} represents a fully connected feedforward network. Naturally these choices are not mandatory and it is customary to use the output of an inner layer to influence layers that precede it, even to the point of altering the structure of the network. An example would be dropping a preceding neuron or entire layer and sending the input through the network again. However here we will not consider these possibilities. It can be proven that a function built out of the neurons in Eq.~(\ref{eq:neurons}), following the procedure outlined in the previous paragraph, can approximate with arbitrary accuracy any continuous function in a compact domain of $\mathbb{R}^N$. For a more precise statement of the relevant theorems I refer to~\cite{Function1989, Hornik1991,Kreinovich1991, Haykin1998}. Here I would like to present a heuristic argument that will also make clear why neural networks provide a good basis for the problem described in the introduction. Take two neurons of the type~(\ref{eq:neurons}) and send their output to a third one. For simplicity consider a one-dimensional input for the two neurons. The function that describes this small neural network is
\ba
f_{NN}(x) =w_1^\prime \sigma(z_1(x))+w_2^\prime \sigma(z_2(x)) + b^\prime\, , \quad z_i(x) = w_i x + b_i\, 
\ea
where $i=1,2$ labels the two initial neurons. For $w_1^\prime=-w_2^\prime=w^\prime$ and $b^\prime=0$ we have
\ba
f_{NN}(x) = w^\prime \left[\sigma(w_1 x + b_1)-\sigma(w_2 x + b_2)\right]\, .
\ea
This is approximately zero for $x\gtrsim -b_2/w_2$ and $x\lesssim -b_1/w_1$ and roughly constant and equal to $w^\prime$ otherwise. By increasing $w_1$ and $w_2$ we can make the transition between zero and $w^\prime$ arbitrarily sharp. By adjusting $b_1$ and $b_2$ we can make the domain over which $f_{NN}(x)$ is non-zero as narrow as we want. So we can make this three-neurons unit generate a smooth peak or a rectangular function. By combining many of these units we can approximate any continuous function as a juxtaposition of rectangular functions. In higher dimensions we can repeat this argument by adding two more neurons for each new direction. We can send all their outputs into a single final neuron and construct a multidimensional rectangular function in the same way. This shows why neural networks are promising candidates for new physics searches. Even if we do not know a priori the type of signal that we are looking for, a network with very few parameters can reproduce an arbitrarily sharp feature, remaining smooth in its absence. This should be contrasted for example with the Fourier transform that would require $\sim 1/\Gamma$ parameters to reproduce a feature of width $\Gamma$. Fewer free parameters mean a smaller look-elsewhere effect and larger sensitivity.
%\begin{figure}[!t]
%\begin{center}
%\includegraphics[width=0.6\textwidth]{sigmoid.pdf}
%\caption{Illustration of how three-neurons with logistic sigmoid activation functions can reproduce a rectangular function or a smooth peak. The parameters in the legend of the plot are defined in Section~\ref{sec:nn}.}
%\label{fig:sigmoid}
%\end{center}
%\end{figure}

To conclude this section I would like to describe very briefly how to determine the values of the free parameters of the network. As anticipated this is not different from fitting using maximum likelihood estimators for the parameters. First one writes down a loss function that is just minus the likelihood and then tries to minimize it. Since the loss functions obtained by nesting neurons are in general non-convex there are no algorithms that are guaranteed to find a global minimum. The prevailing approach consists in finding a ``good enough" local minimum by using Stochastic Gradient Descent. Gradient Descent simply consist in taking a derivative of the loss function and updating the weights and biases by moving a small amount $\epsilon$ in the direction in which the derivative decreases. This technique was proposed by Cauchy in 1847~\cite{S2010}. The parameter $\epsilon$ is called learning rate. It can be fixed a priori or changed adaptively during training. %The celebrated ``backpropagation", used to determine the derivative of the loss function, is nothing more than an application of the chain rule of derivation. 
Since computing the derivative over the entire training sample is usually computationally unfeasible, it is typically computed on a subsample chosen at random (different at every iteration). This is what goes under the name of Stochastic Gradient Descent~\cite{Goodfellow-et-al-2016,Alom2018}. 

Now we have seen how to minimize the loss function, but we still need to discuss how to construct it. As usual the process of building the appropriate likelihood is determined by the specific problem at hand and is best illustrated with an example. Here I discuss what one would do for supervised learning and I refer the reader interested in semi-supervised, unsupervised and reinforcement learning to~\cite{Goodfellow-et-al-2016, Alom2018, bishop:2006:PRML, Hastie2009}. Imagine that you have two sets of pictures one of cats and one of dogs. You would like the network to output $1$ if given a cat and $0$ for a dog. In this case the input $ \vec x$ can be an array of numbers, each representing a different pixel of the picture. Then an obvious choice for the loss function could be
\ba
L[f_{NN}] = \sum_{x \in {\rm cats}} \left[1-f_{NN}(\vec x|\vec w, \vec b) \right]^2+ \sum_{\vec x \in {\rm dogs}} \left[f_{NN}(\vec x|\vec w, \vec b) \right]^2\, . \label{eq:loss}
\ea
At the minimum of $L$,  $f_{NN}(\vec x_{\rm cat})=1$ and $f_{NN}(\vec x_{\rm dog})=0$. %It is very easy to prove it, by taking a functional derivative of $L$ with respect to $f_{NN}$. What is actually implemented in a computer consists in taking the derivatives of $L$ with respect to weights and biases going backwards from the last layer. 
This form of the loss function is just illustrative, in practical applications the cross-entropy, the Kullback-Leibler divergence and their variations are more widely used. One quality that they have over the $\chi^2$ used in~(\ref{eq:loss}) is that their logarithms cancel the exponential saturation of sigmoids and hyperbolic tangents at least for the last layer, making the derivatives larger and the gradient descent algorithm faster for certain values of the input. The process of evaluating $L$ on a subset of the cats and dogs sample, taking its derivatives and updating the values of weights and biases (followed by as many iterations as it takes to obtain an acceptable degree of accuracy) is known as training and the sample used for the process is known as the training sample. The accuracy of the network can be tested on a separate sample, (you guessed it) the testing sample. Clearly this introduction to neural networks is not even remotely comprehensive and several important aspects were not mentioned. For more thorough reviews and books on the subject that I found helpful see~\cite{Goodfellow-et-al-2016, Alom2018, bishop:2006:PRML, Hastie2009}. However we have all the ingredients needed to understand our technique for new physics searches.

\section{Finding New Physics at the LHC}\label{sec:algorithm}
We would like to establish if an experimental dataset follows a theoretical reference model. The situation of interest is one where the dataset is dominated by known processes, i.e. most of the events are described by the reference model, but there can be a statistically significant difference between the probability distribution of the reference model and the one from which the data are extracted. This situation is typical in particle physics and cosmology where standard models have been established and have been extremely successful at describing all the data collected so far. Ideally we would like to find a deviation from these reference models, but a priori we would like to be completely agnostic about the form of this deviation. This is particularly motivated in light of recent results, especially in particle physics. %The two standard models are more successful than ever before, while all their best motivated extensions face increasing difficulties. 

Our technique uses machine learning to approximate the ratio of the probability distribution functions (pdfs) of the data and the reference model. The ratio of pdfs can then be used to construct the Neyman-Pearson test statistic~\cite{10.2307/91247} to determine if the two samples follow the same probability distribution. This is effectively a completely model-independent new physics search.

The domain over which these pdfs are defined depends on our choice of input for the algorithm. It could be all the four vectors of all the particles in a LHC event or, more realistically, a set of observables in a given signal region, where backgrounds and systematics have already been measured using traditional techniques and are under control. For example our data sample could be all LHC events with  at least two muons and the input the four-momenta of the two leading muons. 

To implement this technique, we take a data sample $\mathcal{D}$ of length $N_D$ and a reference sample $R$ of length $\mathcal{N}_R$ (generated using Monte-Carlo) and feed them to a fully connected feedforward network $f_{NN}$ with the following loss function
\ba\label{eq:loss}\displaystyle
L[f]=\sum\limits_{(x,y)}\left[(1-y)\frac{N({\textrm{R}})}{{\mathcal{N}}_{\mathcal{R}}}(e^{f_{NN}(x)}-1)-y\,f_{NN}(x)\right]\,.
\ea
Here $y$ is a label $=1$ for data events and $=0$ for events generated according to the reference model. The variable $x$ represents the space that we are interested to explore (in the above example the four momenta of the two muons). $N(R)$ is the number of expected events in the reference model for the given data sample. It is easy to show that this loss function is proportional to the Neyman-Pearson test statistic:
\ba\label{tNP}
\displaystyle
t({\mathcal{D}})=2\,\log\left[\frac{e^{-N({{\mathbf{\widehat{w}}}})}}{e^{-N({\rm{R}})}}\prod\limits_{x\in {\mathcal{D}}}\frac{n(x|{{\mathbf{\widehat{w}}}})}{n(x|{\rm{R}})}\right]=
-2\,\underset{\{{\mathbf{w}}\}}{\rm{Min}}\left[
N({{\mathbf{{w}}}})-N({\rm{R}})-\sum\limits_{x\in {\mathcal{D}}}f_{NN}(x;{{{\mathbf{w}}}})
\right]
\,,
\ea
where $\mathbf{\widehat w}$ are the parameters of the network at the end of training and $n$ are probability distributions normalized to the total number of events: $n(\cdot | R)$ in the reference model and $n(\cdot | \mathbf{\widehat w})$ as learned by the network for the data. Estimating $N({{\mathbf{{w}}}})$ by the Monte Carlo method
\ba\label{eq:MCint}
\displaystyle
N({{\mathbf{{w}}}})=\frac{N({\textrm{R}})}{{\mathcal{N}}_{\mathcal{R}}}\sum\limits_{x\in{\mathcal{R}}}e^{f(x;{{{\mathbf{w}}}})}\, ,
\ea
we obtain
\ba\label{tML}
\displaystyle
t({\mathcal{D}})=
-2\,\underset{\{{\mathbf{w}}\}}{\rm{Min}}\left[\frac{N({\textrm{R}})}{{\mathcal{N}}_{\mathcal{R}}}\sum\limits_{x\in{\mathcal{R}}}(e^{f(x;{{{\mathbf{w}}}})}-1)-\sum\limits_{x\in {\mathcal{D}}}f(x;{{{\mathbf{w}}}})
\right]\equiv-2\,\underset{\{{\mathbf{w}}\}}{\rm{Min}}\,L[f(\,\cdot\,,{\mathbf{w}})]\, .
\ea

%\be
%L_{\rm ext}[f_{NN}] = \sum_{x\in D} \log[1+e^{-f_{NN}(x)}] +  \frac{\overline{N}_R}{N_R}\sum_{x\in R} \log [1+e^{f_{NN}(x)}]\, ,
%\ee
%where $N(R)$ is the number of expected events in the reference model for the given data sample. The ratio in front of the second term accounts for the fact that the Monte-Carlo sample might have much higher statistics than the number of expected events, but we do not want to render irrelevant the part of the loss function that depends on the data. For simplicity we indicate with the variable $x$ our chosen input, which can of course be multidimensional.

For large enough statistics $\mathcal{N}_R, N_D \gg 1$, it is easy to show that the minimum of the loss function corresponds to
\ba
f_{NN}(x,{{{\mathbf{\widehat{w}}}}})\simeq \log\left[\frac{{{{n}}(x|{\rm{T}})}}{{{{n}}(x|{\rm{R}})}}\right] \, ,
\ea
with $n(x|\rm{T})$ the true data distribution. So the network has learned the ratio between the data and reference model pdfs normalized to the total number of events. Furthermore, the loss function at the end of training is proportional to the Neyman-Person test statistic: $L=- t({\mathcal{D}})/2$.

%the loss function is approximately
%\be
%L_{\rm ext}[f_{NN}] \approx \int dx f_{D}(x)  \log[1+e^{-f_{NN}(x)}] +\int dx f_{R}(x)  \log[1+e^{f_{NN}(x)}]\, ,
%\ee
%where $f_{D}$ and $f_{R}$ are the data and reference model pdfs, respectively. $f_D$ is normalized to the observed ($N_D$) number of events in the sample and $f_R$ to the expected one ($N_R$). Using functional analysis it is easy to conclude that at the minimum of the loss function, $\hat f_{NN}$, the network has learned the log-ratio of pdfs
%\be
%\hat f_{NN}(x)=\log \frac{f_D(x)}{f_R(x)}\, .
%\ee
%As mentioned above we can use this log-ratio to construct the Neyman-Pearson test statistic
%\be
%t(D)= 2 \left(N_R-N_D\right) + 2 \sum_{x\in D} \hat f_{NN}(x) \approx 2 \left(N_R-N_D\right) + 2 \sum_{x\in D} \log \frac{f_D(x)}{f_R(x)}\, .
%\ee
Now we just need to understand if the value of $t(\mathcal{D})$ signals a deviation from the reference model. To do it we need to construct a probability distribution for $t$ in the reference model. This probability distribution has to account for the fact that we have used the data twice, first to learn $f_{NN}(x,{{{\mathbf{\widehat{w}}}}})$ and then to compute $t(\mathcal{D})$. Furthermore it has to correctly account for the look-elsewhere effect. When we construct $t$ by minimizing the loss function we are asking if {\it any} of the non-linear functions parametrized by the neural network can describe the data better than the reference model. Given statistical fluctuations the answer is bound to be yes, unless we correct for the number of hypotheses, alternative to the reference model, that we have tested using the neural network. There is a simple way to solve both problems at once. We can repeat every step of the above procedure substituting the data sample with an instance of the reference model of size $N_D$. So we redo the training using two samples of different sizes ($N_D$ and $\mathcal{N}_R$), but both generated according to the reference model probability distribution. If done multiple times, this is going to give us a list of values of $t$ in the reference hypothesis. From this list we can construct empirically the probability distribution  of $t$ in the reference model, $P(t|R)$. Finally
\ba
p = \int_{t(\mathcal{D})}^{\infty} P(t|R)\, ,
\ea
gives a global $p$-value that we can use to quantify the agreement between the data and the reference model. This solves the problems described above if we fix a neural network a priori rather than using multiple architectures on the same dataset. For methods that allow to select the appropriate neural network I refer to~\cite{DAgnolo:2018cun}.

It is conceptually straightforward to account for systematic uncertainties. We can repeat the procedure outlined above multiple times, after appropriately shifting nuisance parameters, thus obtaining a broader $P(t|R)$ that accounts also for systematics. %We can also be agnostic about normalizations (which would be equivalent to normalizing our expected events in the reference hypothesis to the observed value in the data) by using as a loss function
%\be
%L[f_{NN}] = \frac{1}{N_D}\sum_{x\in D} \log[1+e^{-f_{NN}(x)}] +  \frac{1}{N_R}\sum_{x\in R} \log [1+e^{f_{NN}(x)}]\, .
%\ee
The main limiting factor of this technique is the look-elsewhere effect that the neural network introduces. However the latter can not be avoided in any attempt to search for new physics model-independently and it is not a limitation of the method itself.

Clearly this method has many potential applications.  The possibility of digging in cosmological and collider data in a completely model-independent way is exciting and might offer us surprises also when applied to the datasets that have already been analyzed. Its application to real LHC data is already ongoing inside the CMS collaboration. A more thorough account of this technique, including tests of its performances can be found in~\cite{DAgnolo:2018cun}.

\bibliographystyle{utphys}
\bibliography{bibliography_arxiv}

\providecommand{\href}[2]{#2}\begingroup\raggedright\begin{thebibliography}{10}

\bibitem{DAgnolo:2018cun}
R.~T. D'Agnolo and A.~Wulzer, ``{Learning New Physics from a Machine},''
\href{http://arxiv.org/abs/1806.02350}{{\tt arXiv:1806.02350 [hep-ph]}}.
%%CITATION = ARXIV:1806.02350;%%.

\bibitem{Metodiev:2017vrx}
E.~M. Metodiev, B.~Nachman, and J.~Thaler, ``{Classification without labels:
  Learning from mixed samples in high energy physics},''
  \href{http://dx.doi.org/10.1007/JHEP10(2017)174}{{\em JHEP} {\bf 10} (2017)
  174},
\href{http://arxiv.org/abs/1708.02949}{{\tt arXiv:1708.02949 [hep-ph]}}.
%%CITATION = ARXIV:1708.02949;%%.

\bibitem{Komiske:2018oaa}
P.~T. Komiske, E.~M. Metodiev, B.~Nachman, and M.~D. Schwartz, ``{Learning to
  Classify from Impure Samples},''
\href{http://arxiv.org/abs/1801.10158}{{\tt arXiv:1801.10158 [hep-ph]}}.
%%CITATION = ARXIV:1801.10158;%%.

\bibitem{Dery:2017fap}
L.~M. Dery, B.~Nachman, F.~Rubbo, and A.~Schwartzman, ``{Weakly Supervised
  Classification in High Energy Physics},''
  \href{http://dx.doi.org/10.1007/JHEP05(2017)145}{{\em JHEP} {\bf 05} (2017)
  145},
\href{http://arxiv.org/abs/1702.00414}{{\tt arXiv:1702.00414 [hep-ph]}}.
%%CITATION = ARXIV:1702.00414;%%.

\bibitem{Function1989}
G.~Cybenko, ``{Approximation by Superpositions of a Sigmoidal Function},''
  \href{http://dx.doi.org/10.1007/BF02551275}{{\em Math. Control. Signals,
  Syst.} (1989)  303--314}.

\bibitem{Hornik1991}
K.~Hornik, ``{Approximation Capabilities of Multilayer Feedforward Networks},''
  \href{http://dx.doi.org/http://dx.doi.org/10.1016/0893-6080(91)90009-T}{{\em
  Neural Networks} {\bf 4} (1991) no.~2, 251--257}.

\bibitem{Kreinovich1991}
V.~Y. Kreinovich, ``{Arbitrary nonlinearity is sufficient to represent all
  functions by neural networks: A theorem},''
  \href{http://dx.doi.org/10.1016/0893-6080(91)90074-F}{{\em Neural Networks}
  {\bf 4} (1991) no.~3, 381--383}.

\bibitem{Haykin1998}
S.~Haykin, {\em {Neural Networks: A Comprehensive Foundation}}, vol.~2.
\newblock Prentice Hall, 1998.

\bibitem{Cowan:1998ji}
G.~Cowan, {\em {Statistical data analysis}}.
\newblock Oxford, UK: Clarendon,
1998.
\newblock
%%CITATION = INSPIRE-483505;%%.

\bibitem{S2010}
M.~{Augustine Cauchy}, ``{M{\'{e}}thode g{\'{e}}n{\'{e}}rale pour la
  r{\'{e}}solution des syst{\`{e}}mes d'{\'{e}}quations simultan{\'{e}}es},''
  {\em Comptes Rendus Hebd. S{\'{e}}ances Acad. Sci.} (1847) no.~25, 536--538.

\bibitem{Goodfellow-et-al-2016}
I.~Goodfellow, Y.~Bengio, and A.~Courville, {\em Deep Learning}.
\newblock MIT Press, 2016.
\newblock \url{http://www.deeplearningbook.org}.

\bibitem{Alom2018}
M.~Z. Alom, T.~M. Taha, C.~Yakopcic, S.~Westberg, M.~Hasan, B.~C. {Van Esesn},
  A.~A.~S. Awwal, and V.~K. Asari, ``{The History Began from AlexNet: A
  Comprehensive Survey on Deep Learning Approaches},''
  \href{http://arxiv.org/abs/1803.01164}{{\tt arXiv:1803.01164}}.
  \url{http://arxiv.org/abs/1803.01164}.

\bibitem{bishop:2006:PRML}
C.~M. Bishop, {\em Pattern Recognition and Machine Learning}.
\newblock Springer, 2006.

\bibitem{Hastie2009}
T.~Hastie, R.~Tibshirani, and J.~Friedman,
  \href{http://dx.doi.org/10.1007/b94608}{{\em {The Elements of Statistical
  Learning}}}, vol.~1.
\newblock 2009.
\newblock \href{http://arxiv.org/abs/arXiv:1011.1669v3}{{\tt
  arXiv:arXiv:1011.1669v3}}.
\newblock \url{http://www.springerlink.com/index/10.1007/b94608}.

\bibitem{10.2307/91247}
J.~Neyman and E.~S. Pearson, ``On the problem of the most efficient tests of
  statistical hypotheses,'' {\em Philosophical Transactions of the Royal
  Society of London. Series A, Containing Papers of a Mathematical or Physical
  Character} {\bf 231} (1933)  289--337.
  \url{http://www.jstor.org/stable/91247}.

\end{thebibliography}\endgroup
%\bibliography{bibliography_arxiv}

\end{document}